\documentclass[aps,prl, twocolumn,showpacs]{revtex4}
\usepackage{dcolumn}
\usepackage{bm}
\usepackage[dvips]{graphics}
\begin{document}
\title{High Energy Quantum Teleportation Using Neutral Kaons}
[{\em Published in Phys. Lett. B 641 (2006) 75-80.}]
\author{Yu Shi\footnote{Email address: yushi@fudan.edu.cn}}
\affiliation{Department of Physics, Fudan
University, Shanghai 200433, China}

\begin{abstract}
We describe a scheme of stochastic implementations of quantum
teleportation and entanglement swapping in terms of neutral kaons.
In this scheme, the kaon whose state is to be teleported collides
with one of the two entangled kaons in an Einstein-Podolsky-Rosen
state. Subsequent detection of the outgoing particles of the
collision completes the two-qubit projection on Alice side. There
appear  novel features, which connects quantum information science
with fundamental laws of particle physics.
\end{abstract}


\maketitle

Recent years witnessed the blossom of the subject of quantum
information~\cite{bendiv}. A major topic is quantum teleportation,
the transmission of a quantum state without transporting the
physical object through the intervening space~\cite{ben}, which has
been implemented in optical, atomic and NMR
systems~\cite{teleportation}. Theoretical studies have also been
made in a solid state electron-hole system~\cite{beenakker} and in a
non-inertial frame~\cite{general}. The idea of teleportation is even
involved in studies on black hole evaporation~\cite{horowitz}. In a
closely related process called entanglement swapping, two qubits
which never meet become entangled~\cite{swap}. It  has been
implemented optically~\cite{experiment}.  Virtually all areas of
physics reigned by quantum mechanics have been explored for
possibilities of implementing quantum information processes, but
with an exception: high energy physics. Yet there have been many
explorations on testing Bell theorem in terms of neutral kaons or
$B$-mesons~\cite{bertlmann,leeyang,lot,cern,go}. In this Letter, we
make a theoretical proposal on high energy implementations of
quantum teleportation and entanglement swapping by using neutral
kaons. To our knowledge, it is the first proposal of quantum
information processing in terms of massive elementary particles and
in presence of particle decays. In this novel scheme, the kaon whose
state is to be teleported collides with one of the entangled kaons,
and the detection of the outgoing particles effectively realizes the
two-qubit projection on Alice side, as required in a teleportation
procedure. The projection basis is different from the Bell basis,
but still contains the entangled state same as the original one
shared between Alice and Bob. As a fundamental property of a massive
elementary particle, the teleported degree of freedom, namely, being
$K^0$ or $\bar{K}^0$, and its entanglement, are Lorentz invariant,
in contrast with the case of spin. The cross-fertilization between
quantum information science and high energy physics points to
interesting new directions of research concerning some most
fundamental aspects of information and matter, and may also lead to
useful applications.

The neutral kaon $K^0$ is a meson composed of  quarks $d$ and
$\bar{s}$, while its antiparticle $\bar{K}^0$ is composed of
$\bar{d}$ and $s$~\cite{perkins}. Each of them is a pseudoscalar
with $J^P=0^-$, that is, the angular momentum is  $J=0$, while its
intrinsic parity is negative, i.e.  $P|K^0\rangle = -|K^0\rangle$
and $P|\bar{K}^0\rangle = -|\bar{K}^0\rangle$. They are also
eigenstates of strangeness $S$ with eigenvalues $1$ and $-1$,
respectively, as well as eigenstates of the component $I_3$ of
isospin ($I=1/2$), with eigenvalues $1/2$ and $-1/2$, respectively.
They transform to each other by charge conjugation $C$, i.e.
$C|K^0\rangle = \eta|\bar{K}^0\rangle$ and $C|\bar{K}^0\rangle =
\eta'|K^0\rangle$, where $\eta$ and $\eta'$ are arbitrary phase
factors, and are set to be $-1$ here. Under this convention,
$$|\bar{K}^0\rangle = CP |K^0\rangle,$$
$$|K^0\rangle = CP |\bar{K}^0\rangle.$$
Hence the eigenstates of $CP$ are $$|K_1\rangle =
\frac{1}{\sqrt{2}}(|K^0\rangle + |\bar{K}^0\rangle),$$ with $CP=1$,
and $$|K_{2}\rangle = \frac{1}{\sqrt{2}}(|K^0\rangle
-|\bar{K}^0\rangle),$$  with $CP=-1$. $|K^0\rangle$ and
$|\bar{K}^0\rangle$, or $|K_1\rangle$ and $|K_2\rangle$, expand a
two-dimensional Hilbert space. Another important basis is comprised
of the mass eigenstates $|K_S\rangle$ and $|K_L\rangle$, with
eigenvalues $\lambda_S=m_S-i\Gamma_S/2$ and
$\lambda_L=m_L-i\Gamma_L/2$, where the subscripts ``$S$'' and
``$L$'' stand for the short and long life times of weak decays, with
decay widths $\Gamma_S$ and $\Gamma_L$, respectively. The mass
difference is negligible, as $m_L-m_S = (3.483 \pm 0.006) \times
10^{-12} MeV$, while $m_L \approx m_S \approx m = 497.648 \pm 0.022
MeV $~\cite{data}, hence $m_L-m_S \approx 7.112 \times 10^{-15} m$.
But the mean life times
 $1/\Gamma_S = (0.8953 \pm 0.0006) \times 10^{-10} s$ and
$1/\Gamma_L = (5.114\pm 0.021) \times 10^{-8} s$ differ
significantly~\cite{data}.  In terms of the proper time $\tau$, the
weak decay is described as $$|K_S(\tau)\rangle = e^{-i\lambda_S
\tau} |K_S\rangle,$$
$$|K_L(\tau)\rangle = e^{-i\lambda_L \tau} |K_L\rangle,$$ with $\hbar=c
=1$. $|K_L\rangle$ and $|K_S\rangle$ are related to $CP$ and
strangeness eigenstates  as
\begin{equation}
\begin{array}{lcl}
|K_S\rangle & = & \frac{1}{\sqrt{1+|\epsilon|^2}}(|K_1\rangle
+\epsilon|K_2\rangle) \\ & = &
\frac{1}{\sqrt{|p|^2+|q|^2}}(p|K^0\rangle
+q|\bar{K}^0\rangle), \\
|K_L\rangle & = & \frac{1}{\sqrt{1+|\epsilon|^2}}(|K_2\rangle
+\epsilon|K_1\rangle)\\ & = &
\frac{1}{\sqrt{|p|^2+|q|^2}}(p|K^0\rangle - q|\bar{K}^0\rangle),
\end{array} \label{ksl} \end{equation} where $\epsilon$ is the very
small parameter characterizing $CP$ violation, and is of the order
of $10^{-3}$, $p=1+\epsilon$, $q=1-\epsilon$. In practice, in
neglecting $CP$ violation, one can set $p= q =1$.

Note that the phase factors $\eta$ and $\eta'$ above can be chosen
arbitrarily. If one adopts the convention $\eta=\eta'=1$, then in
Eq.~(\ref{ksl}), the expressions for $|K_S\rangle$ and $|K_L\rangle$
should be exchanged, and in the calculation results below,
$\lambda_S$ and $\lambda_L$ should be exchanged. Anyway,
$|K_S\rangle$ is always dominated by $CP=1$ state,  while
$|K_L\rangle$ is always dominated by $CP=-1$ state.

It has been noted for a long time that from the strong decay of a
vector meson $\phi$ or from the annihilation of a proton-antiproton
pair, a $K^0\bar{K}^0$ pair can be created in  an entangled
Einstein-Podolsky-Rosen (EPR) state~\cite{leeyang}
$$|\Psi_{-}\rangle \equiv
\frac{1}{\sqrt{2}}(|K^0\rangle|\bar{K}^0\rangle
-|\bar{K}^0\rangle|K^0\rangle).$$ $\phi$ mesons can be generated in
electron-positron annihilation with center of mass energy about $1
GeV$, as done in $\phi$ factories. Similar EPR state can be produced
in $B^0\bar{B}^0$ pair from $\Upsilon (4S)$ resonance, which can be
generated in electron-positron annihilation at about $10GeV$, as in
$B$ factories. A lot of discussions were made on how to employ the
entangled $K^0\bar{K}^0$ or $B^0\bar{B}^0$ pair to test nonlocality
or Bell theorem~\cite{bertlmann,leeyang, lot}. Experimentally, EPR
correlation has been confirmed in $K^0\bar{K}^0$ pairs produced in
proton-antiproton annihilation in the CPLEAR detector in
CERN~\cite{cern}, in $K^0\bar{K}^0$ pairs produced in $\phi$ decay
in the KLOE detector in DA$\Phi$NE~\cite{kloe}, as well as in
$B^0\bar{B}^0$ pairs produced in the Belle detector in the KEKB
electron-position collider~\cite{go}.

Our proposal is to let Alice and Bob share an entangled
$K^0\bar{K}^0$ (or $B^0\bar{B}^0$) pair, denoted as $a$ and $b$, and
use it to teleport to $b$ an unknown state of another kaon (or
$B$-meson) $c$, be it in a pure state or entangled with another
system. Our scenario involves collision between kaons $c$ and $a$,
and subsequent measurement of the outgoing particles of $c-a$
collision.

At laboratory time $t=t_y=0$, an entangled $K^0\bar{K}^0$ pair $a$
and $b$ is created as $|\Psi_-\rangle$.  Thus
$$\begin{array}{lcl}
|\Psi_{ab}(0)\rangle  & = &
\frac{1}{\sqrt{2}}(|K^0\rangle_a|\bar{K}^0\rangle_b-
|\bar{K}^0\rangle_a|K^0\rangle_b) \\ & = &
\frac{r}{\sqrt{2}}(|K_L\rangle_a|K_S\rangle_b-
|K_S\rangle_a|K_L\rangle_b), \end{array}$$ where
$r=(|p|^2+|q|^2)/2pq \approx 1$. Up to $r$, the singlet in
strangeness basis is also a singlet in mass basis even though $CP$
violation is taken into account. For convenience, the observers, or
Alice and Bob, which are particle detectors, stay in the laboratory
frame, which is supposed to coincide with the center of mass frame
of $a$ and $b$; the generalization to otherwise case is
straightforward.

After creation, the kaons naturally decay under weak interaction. It
can be found that
$$|\Psi_{ab}(t)\rangle
 = M(t) |\Psi_-\rangle_{ab},$$
where $M(t) = \exp [-i(\lambda_S+\lambda_L)\gamma_{b}t]$. $\gamma_i$
is the Lorentz factor $1/\sqrt{1-v_i^2}$ for particle $i$ with
velocity $v_i$. It has been assumed that $\gamma_a=\gamma_b$.  It
can be estimated that with such decay widths and with the center of
mass energy of about $1 GeV$, the process should be completed within
a few centimeters from the source in the laboratory frame, as indeed
so in the CERN experiment~\cite{cern}.   Interestingly,
$|\Psi_{ab}(t)\rangle $ is Lorentz invariant, as the kaons are
spinless pseudoscalars. The Lorentz invariance of this entanglement
is an advantage over the spin entanglement, which is not Lorentz
invariant in general~\cite{peres2}.

Next we consider the third kaon $c$ generated at time $t_z$ as
$$|\Psi_c(t_z)\rangle = \alpha |K^0\rangle_c +\beta
|\bar{K}^0\rangle_c,$$
which may be unknown. For $t \geq t_z$,
\begin{equation}
|\Psi_c(t)\rangle = F(t) |K^0\rangle_c +G(t) |\bar{K}^0\rangle_c,
\label{ct}
\end{equation}
with $F(t)=[(\alpha+\beta p/q)e^{-i\lambda_S\gamma_c
(t-t_z)}+(\alpha-\beta p/q) e^{-i\lambda_L \gamma_c (t-t_z)}]/2$,
$G(t)=[(\alpha q/p+\beta )e^{-i\lambda_S \gamma_c (t-t_z)}-(\alpha
q/p -\beta ) e^{-i\lambda_L \gamma_c (t-t_z)}]/2$.

The state of the three particles is thus
$$|\Psi_{cab}(t)\rangle
=|\Psi_c(t)\rangle\otimes|\Psi_{ab}(t)\rangle.$$

Consider the following basis states of $c-a$, which are eigenstates
of $P$, $S$ and $I$: $|\phi_1\rangle_{ca} \equiv |K^0K^0\rangle$
with $P=1$, $S=2$, $I=1$; $|\phi_2\rangle_{ca}
\equiv|\bar{K}^0\bar{K}^0\rangle$ with $P=1$, $S=-2$, $I=1$;
$|\phi_3\rangle_{ca} \equiv |\Psi_+\rangle \equiv
\frac{1}{\sqrt{2}}(|K^0\rangle|\bar{K}^0\rangle +
|\bar{K}^0\rangle_c|K^0\rangle)$,  with $P=1$, $S=0$, $I=1$;  and
$|\phi_4\rangle_{ca} \equiv|\Psi_-\rangle$ with $P=-1$, $S=0$,
$I=0$. For the reason which will be clear shortly, we rewrite
$|\Psi_{cab}(t)\rangle$ in terms of these four basis states of $c-a$
as
\begin{equation}
\begin{array}{lll} |\Psi_{cab}(t)\rangle & = &
\frac{M(t)}{2}\{\sqrt{2}F(t)|\phi_1\rangle_{ca}|\bar{K}^0\rangle_b
\\&&-\sqrt{2}G(t)|\phi_2\rangle_{ca}|K^0\rangle_b \\
&& -|\phi_3\rangle_{ca}[F(t)|K^0\rangle_b-G(t)|\bar{K}^0\rangle_b] \\
&&
-|\phi_4\rangle_{ca}[F(t)|K^0\rangle_b+G(t)|\bar{K}^0\rangle_b]\}.
\end{array}
\label{en2}
\end{equation}

We design the set-up in such a way that $a$ and $c$ fly in opposite
directions and towards each other, hence they destine to collide at
a certain position $x$ at a certain time $t_{x}$ (FIG.~1 shows the
scheme, FIG.~2 is the spacetime  diagram).

\begin{figure}

\resizebox{4cm}{1cm}{
\includegraphics{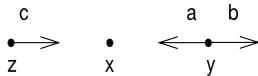}}
\caption{\label{fig1} Scheme of quantum teleportation using an
entangled kaon pair $a$ and $b$ generated from a source $y$. Another
kaon $c$ comes from a source $z$. $a$ and $c$ fly collinearly and
towards each other, thus  collide at a certain time $t_x$ at a
position $x$. }
\end{figure}

\begin{figure}
\resizebox{4cm}{3cm}{\includegraphics[150pt,300pt][425pt,525pt]{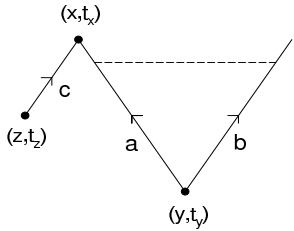}}
\caption{\label{fig2} Spacetime diagram of the kaon teleportation.
The horizontal direction represents the position while the upward
direction represents time flow. The broken line represents the
entanglement. $(z,t_z)$, $(y,t_y)$ and  $(x,t_x)$ are spacetime
coordinates of the generation of $c$, the generation of $a-b$
entangled pair, and the $c-a$ collision, respectively, in the
laboratory frame.}
\end{figure}

Upon collision, $c$ and $a$ become an interacting whole.  The effect
of collision can be represented as a unitary transformation ${\cal
S}$ on $c-a$. The brief and negligible time duration $\delta$ of the
collision is much shorter than the life times of weak decay, thus we
ignore the decay of kaon $b$ during the negligible interval of $c-a$
collision. Therefore, through $c-a$ collision, the state of the
three kaons becomes
\begin{equation}
\begin{array}{lll} |\Psi_{cab}(t_x+\delta)\rangle & = &
\frac{M(t_x)}{2}\{\sqrt{2}F(t_x) {\cal
S}|\phi_1\rangle_{ca}|\bar{K}^0\rangle_b \\
&&-\sqrt{2}G(t_x) {\cal
S}|\phi_2\rangle_{ca}|K^0\rangle_b \\
&& - {\cal
S}|\phi_3\rangle_{ca}[F(t_x)|K^0\rangle_b-G(t_x)|\bar{K}^0\rangle_b] \\
&& - {\cal
S}|\phi_4\rangle_{ca}[F(t_x)|K^0\rangle_b+G(t_x)|\bar{K}^0\rangle_b]\}.
\end{array}
\label{en22}
\end{equation}

$\{{\cal S}|\phi_i\rangle_{ca}\}$ is also a basis of $c-a$ system.
Furthermore, as $c-a$ collision is governed by strong interaction,
${\cal S}$ conserves $S$, $P$ and $I$, thus ${\cal
S}|\phi_i\rangle_{ca}$ ($i=1,2,3,4$) is still an eigenstate of $S$,
$P$ and $I$, with the same eigenvalues as those for
$|\phi_i\rangle_{ca}$.

The two-particle projection of $c-a$ is completed when the outgoing
particles of $c-a$ collision are detected. The particle detector,
close and around the collision point, playing the role of Alice,
detects particles with specific values of $S$, $P$ and $I$, thus
will have projected $c-a$ to one of the four eigenstates ${\cal
S}|\phi_i\rangle_{ca}$. Note that it may  be particles other than
kaons that are detected in ${\cal S}|\phi_i\rangle_{ca}$. But this
does not matter, as $S$, $P$ and $I$ are conserved by ${\cal S}$.
The detector may be based on strong interaction with absorbers, and
constructed in a way similar to that in CPLEAR~\cite{cern}.

According to standard quantum  theory, upon measurement (detection),
the state instantaneously projects to an eigenstate of the
observable~\cite{dirac}. The reference frame can be chosen
arbitrarily, but once it is chosen, the instantaneous projection
needs to be consistently made~\cite{peres}.

The probability of projection to ${\cal S}|\phi_i\rangle_{ca}$ is
calculated as $\langle\Psi_{cab}(t_x+\delta) {\cal
S}|\phi_i\rangle\langle\phi_i|{\cal
S}^{\dagger}\Psi_{cab}(t_x+\delta)\rangle$. It is calculated that
for $i=1,2,3,4$, the probabilities are $|M(t_x)|^2|F(t_x)|^2/2$,
$|M(t_x)|^2|G(t_x)|^2/2$, $|M(t_x)|^2[|F(t_x)|^2+|G(t_x)|^2]/4$, and
$|M(t_x)|^2[|F(t_x)|^2+|G(t_x)|^2]/4$, respectively.

Thus the two-qubit operation by Alice in  the teleportation protocol
is effectively realized. Conditioned on $P$, $S$  and $I$ of the
detected outgoing particles of $c-a$ collision, the state of $b$ is
known to be correspondingly in one of the four states
$|\bar{K}^0\rangle_b$, $|K^0\rangle_b$,
$[F(t_x)|K^0\rangle_b-G(t_x)|\bar{K}^0\rangle_b]/\sqrt{|F(t_x)|^2+|G(t_x)|^2}$,
$[F(t_x)|K^0\rangle_b+G(t_x)|\bar{K}^0\rangle_b]/\sqrt{|F(t_x)|^2+|G(t_x)|^2}$.
A noteworthy point is that despite the decay, the state after
projection should still be normalized; the decay effect has been
taken into account in the projection probability. As $P$ and $S$ are
already sufficient to distinguish ${\cal S}|\phi_i\rangle_{ca}$ for
different $i$'s, it is not necessary to consider $I$.

It is difficult to implement subsequent precise one-bit unitary
transformations on $b$ particle, which is a part of the conventional
scheme of teleportation. Hence we suggest to adopt a stochastic
strategy, as follows. With negligible time delay, upon receiving the
communication of the projection result of Alice, Bob decides whether
to retain or abandon $b$ particle according to whether the state is
what he needs. In actual experiments, the classical communication
and the subsequent conditional operation can be realized by an
automatic control system. As in the usual scenario of teleportation,
suppose Bob wishes to obtain
$[F(t_x)|K^0\rangle_b+G(t_x)|\bar{K}^0\rangle_b]/\sqrt{|F(t_x)|^2+|G(t_x)|^2}$.
According to the expansion in Eq.~(\ref{en22}), Bob should retain
the particle if and only if the projection result of $c$ and $a$ is
${\cal S}|\Psi_-\rangle_{ca}$.

The four possible projection results at $t_x+\delta$ lead to
different values of strangeness ratio $\xi(t \geq t_x+\delta)$ of
$b$ particle, which can be experimentally verified in terms of
reaction with nuclear matter. For $|\Psi(t \geq t_x+\delta)\rangle_b
= f(t) |K^0\rangle_b + g(t)|\bar{K}^0\rangle_b$, $\xi(t)\equiv
|f(t)|^2/g(t)|^2$. Many runs of the procedure are needed to measure
this quantity. If irrespective of the projection results of $c-a$,
$b$ particles in the different runs of the experiment are all
considered in measuring $\xi(t)$, then $\xi(t)$ should be calculated
by using $|\Psi_{cab}(t)\rangle$, consequently $\xi(t)=1$. In
contrast, if only $b$ particles in those runs of the experiment with
a certain projection result of $c-a$ are considered in measuring
$\xi(t)$, then $\xi(t)$ is calculated by using the corresponding
projected state of $b$. Denote the state of $b$ following the
projection as $\alpha(t_x)|K^0\rangle+\beta(t_x)|\bar{K}^0\rangle$.
Its subsequent evolution is then similar to Eq.~(\ref{ct}), with
$t_z$ substituted by $t_x+\delta$, $\gamma_c$ by $\gamma_b$,
$\alpha$ by $\alpha(t_x)$, $\beta$ by $\beta (t_x)$. It can be found
that $\xi(t)=|\alpha(t_x)(e^{-\Gamma_S\tau/2}+e^{-\Gamma_L\tau/2})+
\beta(t_x)(e^{-\Gamma_S\tau/2}-e^{-\Gamma_L\tau/2})|^2/
|\alpha(t_x)(e^{-\Gamma_S\tau/2}-e^{-\Gamma_L\tau/2})+
\beta(t_x)(e^{-\Gamma_S\tau/2}+e^{-\Gamma_L\tau/2})|^2$, where
$\tau=\gamma_b(t-t_x-\delta)$. For each of the four projection
cases, $\xi(t)$ is very different from $1$. For example, if the
teleportation is successful, i.e. the projection result of $c-a$ is
$|\Psi_-\rangle_{ca}$, then $\xi(t)$ is given by substituting
$\alpha(t_x)=F(t_x)/\sqrt{|F(t_x)|^2+|G(t_x)|^2}$,
$\beta(t_x)=G(t_x)/\sqrt{|F(t_x)|^2+|G(t_x)|^2}$. Like delay choice
in entanglement swapping~\cite{peres4}, the projection results of
$c-a$ can even be revealed  only after all the experiments are
finished, and are then used to sort the runs of the procedure to
four subensembles corresponding to the four projection results.

Now we consider a similar stochastic implementation of entanglement
swapping. In addition to $|\Psi_-\rangle_{ab}$ generated at time
$t_y=0$, another kaon pair $d$ and $c$ is generated as
$|\Psi_-\rangle_{dc}$ at time $t_z$. Similar to
$|\Psi_{ab}(t)\rangle$, we have
$$|\Psi_{dc}(t)\rangle = M'(t-t_z)|\Psi_-\rangle_{dc},$$
where $M'(t-t_z) = \exp [-i(\lambda_S+\lambda_L)\gamma_{d}(t-t_z)]$,
supposing $\gamma_c=\gamma_d$.
Thus the state of the four particles is
$$|\Psi_{dcab}(t)\rangle =
M'(t-t_z)M(t) |\Psi_-\rangle_{dc}|\Psi_-\rangle_{ab}.$$ In terms of
the $c-a$ eigenstates of $P$, $S$ and $I$, as given above,
$|\Psi_{dcab}(t)\rangle$ can be written as
$$\begin{array}{lll}
|\Psi_{dcab}(t)\rangle &
=&\frac{M'(t-t_z)M(t)}{2}(|\Psi_+\rangle_{ca}|\Psi_+\rangle_{db}-
|\Psi_-\rangle_{ca}|\Psi_-\rangle_{db} \\ &&
-|K^0K^0\rangle_{ca}|\bar{K}^0\bar{K}^0\rangle_{db}-
|\bar{K}^0\bar{K}^0\rangle_{ca}|K^0K^0\rangle_{db}).\end{array}$$

Similar to the above scenario of pure state teleportation, we let
$c$ and $a$ fly towards each other to collide at a certain time
$t_x$ (FIG.~3 and FIG.~4). Within a negligible time interval
$\delta$, the collision, effecting a unitary transformation ${\cal
S}$ on $c-a$, evolves $|\Psi_{dcab}(t_x)\rangle$ to
$$\begin{array}{lll}
|\Psi_{dcab}(t_x+\delta)\rangle &
=&\frac{M'(t_x-t_z)M(t_x)}{2}({\cal
S}|\Psi_+\rangle_{ca}|\Psi_+\rangle_{db} \\&& - {\cal
S}|\Psi_-\rangle_{ca}|\Psi_-\rangle_{db} \\ && -{\cal
S}|K^0K^0\rangle_{ca}|\bar{K}^0\bar{K}^0\rangle_{db}\\&&- {\cal
S}|\bar{K}^0\bar{K}^0\rangle_{ca}|K^0K^0\rangle_{db}).\end{array}$$
Then, in detecting outgoing particles from $c-a$ collision, $c$ and
$a$ are projected to one of the four states ${\cal
S}|\Psi_{+}\rangle_{ca}$, ${\cal S}|\Psi_-\rangle_{ca}$, ${\cal
S}|K^0K^0\rangle_{ca}$ and ${\cal
S}|\bar{K}^0\bar{K}^0\rangle_{ca}$, and then $P$, $S$ and $I$ are
measured. Correspondingly $d$ and $b$ are projected to
$|\Psi_{+}\rangle_{ca}$, $|\Psi_-\rangle_{ca}$,
$|K^0K^0\rangle_{ca}$ and $ |\bar{K}^0\bar{K}^0\rangle_{ca}$,
respectively, each with probability $|M'(t_x-t_z)M(t_x)|^2/4$.
Again, the projection result is revealed by $P$, $S$ and $I$ of the
outcomes of $c-a$ collision, according to which Bob chooses to
retain or abandon $b$ particle.

The effect of entanglement swapping can be verified by measuring the
strangeness asymmetry  between $b$ and $d$, defined as
$A(t)=[p_{diff}(t)-p_{same}(t)]/[p_{diff}(t)+p_{same}(t)]$, where
$p_{diff}(t)$ and $p_{same}(t)$ are, respectively, the probabilities
for $b$ and $d$ to have different and same strangeness
values~\cite{cern}. Many runs of the experiment are needed to
experimentally determine $A(t)$. If all the $d-b$ pairs in different
runs are considered, irrespective of the projection results of
$c-a$, then $A(t)=0$, as calculated from $|\Psi_{dcab}(t)\rangle$.
In contrast, if only the $d-b$ pairs corresponding to a certain
projection result  of $c-a$ collision are considered,  then $A(t)$
is calculated by using the corresponding projected state of $d$ and
$b$. For example, if at $t=t_x+\delta$, $c$ and $a$ are projected to
${\cal S}|\Psi_-\rangle_{ca}$, i.e. the entanglement swapping is
successful, then for $t \geq t_x+\delta$, $|\Psi(t\geq
t_x+\delta)\rangle_{db} = |\Psi(t\geq t_x+\delta)\rangle_{db} =
g_1|K^0\rangle_d|K^0\rangle_b + g_2|K^0\rangle_d|\bar{K}^0\rangle_b
+g_3|\bar{K}^0\rangle_d|K^0\rangle_b
+g_4|\bar{K}^0\rangle_d|\bar{K}^0\rangle_b$, where
$g_1=(e^{-i(\lambda_L\tau_d+\lambda_S\tau_b)}-e^{-i(\lambda_S\tau_d+\lambda_L\tau_b)})p/2\sqrt{2}q$,
$g_4=-(e^{-i(\lambda_L\tau_d+\lambda_S\tau_b)}-e^{-i(\lambda_S\tau_d+\lambda_L\tau_b)})q/2\sqrt{2}p$,
$g_2
=(e^{-i(\lambda_L\tau_d+\lambda_S\tau_b)}+e^{-i(\lambda_S\tau_d+\lambda_L\tau_b)})/2\sqrt{2}$,
$g_3=-g_2$, where $\tau_d=\gamma_d(t-t_x-\delta)$,
$\tau_b=\gamma_b(t-t_x-\delta)$. Consequently $A(t \geq
t_x+\delta)\approx
2e^{-(\Gamma_S+\Gamma_L)(\tau_d+\tau_b)/2}/(e^{-(\Gamma_L\tau_d+\Gamma_S\tau_b)}+
e^{-(\Gamma_L\tau_d+\Gamma_S\tau_b)})$. If at $t=t_x+\delta$, $c$
and $a$ are projected to ${\cal S}|\Psi_+\rangle_{ca}$, then
$|\Psi(t\geq t_x+\delta)\rangle_{db} = g_1|K^0\rangle_d|K^0\rangle_b
+ g_2|K^0\rangle_d|\bar{K}^0\rangle_b
+g_3|\bar{K}^0\rangle_d|K^0\rangle_b
+g_4|\bar{K}^0\rangle_d|\bar{K}^0\rangle_b$, where
$g_1=(e^{-i\lambda_S(\tau_d+\tau_b)}-e^{-i\lambda_L(\tau_d+\tau_b)})p/2\sqrt{2}q$,
$g_4=(e^{-i\lambda_S(\tau_d+\tau_b)}-e^{-i\lambda_L(\tau_d+\tau_b)})q/2\sqrt{2}p$,
$g_2=g_3
=(e^{-i\lambda_S(\tau_d+\tau_b)}+e^{-i\lambda_L(\tau_d+\tau_b)})/2\sqrt{2}$.
Hence $A(t\geq t_x+\delta) \approx
2e^{-(\Gamma_S+\Gamma_L)(\tau_d+\tau_b)/2}/(e^{-\Gamma_S(\tau_d+\tau_b)}+
e^{-\Gamma_L(\tau_d+\tau_b)})$. It can also be calculated that for
$c-a$ projection to ${\cal S}|K^0K^0\rangle_{ca}$ or ${\cal
S}|\bar{K}^0\bar{K}^0\rangle_{ca}$ at $t_x+\delta$,  $A(t \geq
t_x+\delta) \approx -4
e^{-(\Gamma_S+\Gamma_L)\tau_b}/(e^{-2\Gamma_S\tau_b}+
2e^{-(\Gamma_S+\Gamma_L)\tau_b}+ e^{-2\Gamma_L\tau_b})$ in case
$\gamma_b=\gamma_d$; the expressions for the case $\gamma_b\neq
\gamma_d$ are too cumbersome to be included here.  Again, the
projection results of $c-a$ can even be revealed  only after all the
runs of the experiment are finished, and are then used to sort the
runs to four subensembles corresponding to the four projection
results.

\begin{figure}
\resizebox{4cm}{1cm}{\includegraphics{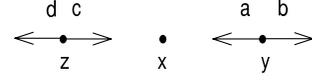}}
\caption{\label{fig3} Scheme of entanglement swapping using
entangled kaon pair $a-b$ generated from a source $y$, and another
entangled kaon pair $c-d$ generated from a source $x$, both in EPR
states. $a$ and $c$ fly collinearly and towards each other, thus
collide at a certain time $t_x$ at a position $x$. }
\end{figure}
\begin{figure}
\resizebox{4cm}{3cm}{\includegraphics[150pt,300pt][425pt,525pt]{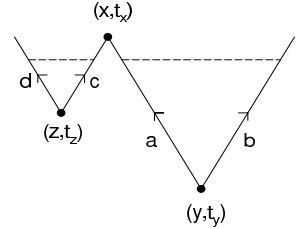}}
\caption{\label{fig4} Spacetime diagram of the entanglement swapping
of kaons.  }
\end{figure}

As noted in the original paper on teleportation~\cite{ben},
teleportation can be made on a qubit $c$ arbitrarily entangled with
any other system $d$, with both the usual teleportation of a pure
state and the entanglement swapping being special cases. The detail
is the following. Suppose an arbitrary unknown state
$|\Psi_0\rangle$ is shared between the qubit $c$ and another system
$d$. One can always write
$|\Psi_0\rangle_{dc} = \gamma |w_{1}\rangle_d \uparrow_c + \rho
|w_{2}\rangle_d \downarrow_c,$  where $|w_{1}\rangle_d$ and
$|w_{2}\rangle_d$ are normalized, but not necessarily orthogonal.
Suppose $a$ and $b$ share a Bell state, say $|\Psi_-\rangle_{ab}$.
Then $|\Psi\rangle_{dcab}$ can be written as
 $$\begin{array}{lll}
 |\Psi\rangle_{dcab} & = & \frac{1}{2}|\Phi_+\rangle_{ca} (\gamma
|w_1\rangle_d \downarrow_b - \rho|w_{2}\rangle_d\uparrow_b) \\
&& +\frac{1}{2}|\Phi_-\rangle_{ca} (\gamma |w_1\rangle_d
\downarrow_b + \rho|w_{2}\rangle_d\uparrow_b)
\\&&-\frac{1}{2}|\Psi_+\rangle_{ca}(\gamma |w_1\rangle_d \uparrow_b -
\rho|w_{2}\rangle_d\downarrow_b)
\\&&-\frac{1}{2}|\Psi_-\rangle_{ca}(\gamma |w_1\rangle_d \uparrow_b +
\rho|w_{2}\rangle_d\downarrow_b).\end{array}$$  A Bell measurement
is performed on $c$ and $a$, with the resulting state
$|\Phi_+\rangle_{ca}$, $|\Phi_-\rangle_{ca}$, $|\Psi_+\rangle_{ca}$
or $|\Psi_-\rangle_{ca}$. Correspondingly, the state of $d-b$
projects to $-i(\sigma_y)_b|\Psi_0\rangle_{db}$,
$(\sigma_x)_b|\Psi_0\rangle_{db}$, $(\sigma_z)_b|\Psi_0\rangle_{db}$
and $|\Psi_0\rangle_{db}$, respectively, where $(\sigma_i)_b$
represents Pauli operation acting on $b$. Therefore,  depending on
the measurement result of  $e-a$, Bob can correspondingly perform on
$b$ operation $\sigma_y$ or $\sigma_x$ or $\sigma_z$ or make no
operation. $|\Psi_0\rangle$ is then teleported to the same state
shared between $d$ and $b$.

Such a general teleportation can also be implemented in high energy
mesons, in a way similar to the above scheme. To prepare an
arbitrarily entangled kaon pair, one can, for example, let one or
both particles of the kaon pair $c-d$, generated as the EPR state
$|\Psi_-\rangle_{dc}$, pass through regeneration materials, which
change the superposition coefficients~\cite{perkins}. Thus one
obtains an unknown state, which is supposed to decay to
$|\Psi_0(t_x)\rangle_{dc} = \gamma |w_{1}\rangle_d |K^0\rangle_c +
\rho |w_{2}\rangle_d |\bar{K}^0\rangle_c$ at time $t_x$ when $c$ and
$a$ collide. With kaon pair $a-b$ prepared at $t=0$ in
$|\Psi_-\rangle_{ab}$, one can write
$$\begin{array}{lll}|\Psi_{dcab}(t_x)\rangle
 & = & \frac{M(t_x)}{2} [\sqrt{2}\gamma|K^0K^0\rangle_{ca}
|w_1\rangle_d|\bar{K}^0\rangle_b \\&&-
\sqrt{2}\rho|\bar{K}^0\bar{K}^0\rangle_{ca}|w_2\rangle_d|K^0\rangle_b\\
&&-|\Psi_+\rangle_{ca}(|w_1\rangle_d |K^0\rangle_b -
\rho|w_2\rangle_d|\bar{K}^0\rangle_b) \\&&-
|\Psi_-\rangle_{ca}(\gamma |w_1\rangle_d |K^0\rangle_b +
\rho|w_2\rangle_d|\bar{K}^0\rangle_b)].\end{array}$$ Afterwards, one
proceeds in the same way as the above scheme of entanglement
swapping.  One can choose to retain the $d-b$ pair only if the
detected outgoing particles of $c-a$ collision are with $P=-1$,
$S=0$ and $I=0$, i.e. $c-a$ is projected to ${\cal
S}|\Psi_-\rangle_{ca}$. In this way, one obtains $\gamma
|w_{1}\rangle_d |K^0\rangle_b + \rho |w_{2}\rangle_d
|\bar{K}^0\rangle_b$.

We see no particular obstacles in actually implementing our
proposal. Especially, with the previous experimental experiences in
nonlocality study~\cite{cern,kloe,go}, it looks feasible to
implement teleportation in these places. Compared with the
experiments on detecting $CP$ violation and on testing Bell theorem,
we only need the additional facilities of realizing kaon collision
and the subsequent detection of the outgoing particles. We do not
need either the precision as high as that in detecting $CP$
violation or the particular arrangements required by the subtle
argument of Bell theorem. Hence in some aspects, implementing the
teleportation scheme here may be easier than detecting $CP$
violation and testing Bell theorem. It seems easier to implement
entanglement swapping than teleportation of either a pure kaon state
or a kaon entangled with another kaon in an unknown state, as only
two pairs of kaons in $|\Psi_-\rangle$ need to be prepared for
entanglement swapping.

It is interesting to study high energy processes involving a single
copy of particles as employed in quantum information protocols, such
that the quantum nature of the processes can be more manifested. On
the other hand, present high energy experiments often employ a beam
of particles consisting of a group of particles prepared in the same
state. Hence many runs of the same procedure mentioned in the above
discussions can actually be done altogether simultaneously, as in
the CPLEAR experiment~\cite{cern}. In detecting $c-a$ collision  and
in analyzing the correlation between $b$ and $d$ particles, one
needs to find the correspondence between events of the entangled
particles in a same copy of state. This can be achieved by analyzing
the particle trajectories and momenta, as done in CPLEAR experiment.
The details of the collision between the kaons and the subsequent
detection need further studies. Moreover, high energy processes in
presence of entanglement with distant particles, as depicted in our
spacetime diagrams, pose a new subject worth detailed
investigations.

To summarize, we have described a scheme of stochastic
implementations of quantum teleportation and entanglement swapping
using neutral kaons.  This work connects quantum information science
to particle physics. The neutral kaon whose state is to be
teleported collides with a neutral kaon which is entangled with
another one in an EPR state.  The detection of the outgoing
particles of the collision completes the projection on Alice's side
to an eigenstate of $P$, $S$ and $I$, which is conserved in the
collision, as governed by strong interaction. Conditioned on this
projection, teleportation or entanglement swapping can be made
stochastically. We also envisage verification schemes based on
strangeness measurements.  We expect our discussion to open up
researches on high energy quantum information, which can stimulate
particle physics to study processes involving entanglement,
projection and decoherence, like similar developments in other areas
of physics. They also furnish a fruitful playground for the
extension of the notions of quantum information science to regimes
of relativity and high
energy~\cite{general,horowitz,peres,peres2,alsing2}. There are some
profound implications. For instance, in conventional teleportation,
it is only the state, rather than the particle which carries the
state, that is teleported. In high energy physics, the particle
itself, e.g. $K^0$ or $\bar{K}^0$, represents a state of the quantum
field, hence can be teleported, as demonstrated here in the example
of neutral kaons.

I thank Professors R. A. Bertlmann, R. Jozsa, D. H. Perkins, S.
Popescu and M. Stone for useful correspondences. This work is
dedicated to the $50th$ anniversary of the resolution of the
$\theta-\tau$ puzzle.

\end{document}